\documentclass[prb,twocolumn,superscriptaddress,amsmath,amssymb,showpacs,floatfix,preprintnumbers,keywords]{revtex4-2}
\usepackage{amsmath}
\usepackage{amssymb}
\usepackage{graphicx}
\usepackage[export]{adjustbox}
\usepackage{epstopdf}
\usepackage{color}
\usepackage{epsfig}
\usepackage{braket}
\usepackage{amsmath}
\usepackage{stackengine}
\usepackage[colorlinks=true,citecolor=blue,linkcolor=blue]{hyperref}
\usepackage{amssymb}

\DeclareMathAlphabet{\bi}{OML}{cmm}{b}{it}

\begin{document}

\title{Atomic collapse in gapped graphene: lattice and valley effects}

\author{Jing Wang}
\email[]{wangjing@hdu.edu.cn}
\affiliation{Key Laboratory of Micro-nano Sensing and IoT of Wenzhou, Wenzhou Institute of Hangzhou Dianzi University, Wenzhou, 325038, China}
\affiliation{School of Electronics and Information, Hangzhou Dianzi University, Hangzhou, Zhejiang Province 310038, China}
\affiliation{Departement Fysica, Universiteit Antwerpen, Groenenborgerlaan 171, B-2020 Antwerpen, Belgium}
\affiliation{NANOlab Center of Excellence, University of Antwerp, Groenenborgerlaan 171, B-2020 Antwerpen, Belgium}

\author{Xiaotai Wu}
\affiliation{School of Electronics and Information, Hangzhou Dianzi University, Hangzhou, Zhejiang Province 310038, China}

\author{Wen-Sheng Zhao}
\affiliation{School of Electronics and Information, Hangzhou Dianzi University, Hangzhou, Zhejiang Province 310038, China}

\author{Yuhua Cheng}
\affiliation{School of Electronics and Information, Hangzhou Dianzi University, Hangzhou, Zhejiang Province 310038, China}
\affiliation{Key Laboratory of Micro-nano Sensing and IoT of Wenzhou, Wenzhou Institute of Hangzhou Dianzi University, Wenzhou, 325038, China}

\author{Yue Hu}
\affiliation{School of Electronics and Information, Hangzhou Dianzi University, Hangzhou, Zhejiang Province 310038, China}
\affiliation{Key Laboratory of Micro-nano Sensing and IoT of Wenzhou, Wenzhou Institute of Hangzhou Dianzi University, Wenzhou, 325038, China}


\author{Fran\c{c}ois M. Peeters}
\email[]{francois.peeters@uantwerpen.be}
\affiliation{Departamento de F\'\i sica, Universidade Federal do Cear\'a, Campus do Pici, Fortaleza, Cear\'a, Brazil}
\affiliation{Departement Fysica, Universiteit Antwerpen, Groenenborgerlaan 171, B-2020 Antwerpen, Belgium}
\affiliation{NANOlab Center of Excellence, University of Antwerp, Groenenborgerlaan 171, B-2020 Antwerpen, Belgium}


\begin{abstract}
We study the atomic collapse phenomenon in $K$ and $K'$ valley of gapped graphene. Bound states induced by Coulomb impurity in the gap turn into atomic collapse resonances as the charge increases beyond the supercritical charge $Z_c$. $Z_c$ increases sublinear with the band gap $\Delta$. The atomic collapse resonances result in peaks in the LDOS at the same energies in $K$ and $K'$ valley, but the strong (weak) LDOS peaks in $K$ valley are weak (strong) LDOS peaks in $K'$ valley reminiscent of pseudospin polarization phenomenon. From a spatial LDOS analysis of the atomic collapse resonance states, we assign specific atomic orbitals to the atomic collapse resonances. Remarkably, the two $p$ atomic orbital atomic collapse states are no longer degenerate and splits into two having lobes in different directions in the graphene plane.
\end{abstract}

\maketitle
\section{Introduction}\label{sec:1}
Twenty years have passed since graphene was first discovered in 2004~\cite{ref1} resulting in a surge of research on graphene and other atomic thin materials. Graphene exhibits an unusual linear dispersion relation around the two valleys situated at $K$ and $K'$ Dirac points. The charge carriers are governed by Dirac-Weyl equation~\cite{ref2}. Graphene has become a  model system to study fundamental physical effect in 2D quantum electrodynamics (QED), such as Klein tunneling, anomalous integer quantum Hall effect and so on. 

Atomic collapse is a relativistic phenomenon in QED that was predicted in 1930~\cite{ref3}. When the nucleus charge $Z$ exceeds a critical value $Z_c\approx170$~\cite{ref4,ref5,ref6}, its surrounding electron spirals down into the nucleus, after which it eject its antimatter opposite, a positron, which would spiral outward. This process is called atomic collapse. As the largest known atomic nuclei in nature is 118, atomic collapse has never been observed in real atoms. What makes graphene interesting, is that the effective critical charge reduces as low as $Z_c \sim 1-2$. The reasion is that the Fermi velocity of graphene is $v_f\approx10^6m/s$ which is 300 times smaller than the velocity of light and consequenctly its effective fine structure constant increases to $\alpha=e^2/(\hbar v_F \kappa)\approx 2.19/\kappa$, where the effective dielectric constant $\kappa$ takes the value between 3 and 10~\cite{ref7,ref8,ref9}. Thus the condition for $Z>1/\alpha$ is easier fulfilled in graphene.

Atomic collapse in graphene was first demonstrated experimentally by the Michael F. Crommie's group~\cite{ref10}. Charged calcium atoms were deposited on the surface of a graphene field-effect transistor and were induced to form "dimers". The dimers were manipulated into “artificial nuclei” using the tip of a scanning tunneling microscope (STM). As the charge of “artificial nuclei” increases from subcritical to supercrticial, an atomic collapse state was detected by STM spectroscopy. Later, atomic collapse was reported in a charged vacancy graphene system~\cite{ref11} and in a nanometer-scale p-n junction in graphene~\cite{ref12}. A new physical region of frustrated supercriticality was found in graphene containing two Coulomb centers~\cite{ref13,ref14,ref15}.

Atomic collapse in graphene was theoretically extended to other systems. Refs.~\cite{ref16,ref17,ref18,ref19,ref20,ref21} studied the atomic collapse in graphene in a single charged impurity field and Refs.~\cite{ref22,ref23,ref24,ref25,ref26,ref27} extended the system to include two identical impurity charges. A strong magnetic field influences the critical charge, but there is no uniform conclusion on how the critical charge changes with the magnetic field ~\cite{ref28,ref29,ref30,ref31,ref32,ref33,ref34,ref35,ref36,ref37}. Conversely, studies consistently show that a band gap in graphene increases the critical charge~\cite{ref38,ref39}. Band gap is an important parameter in graphene semiconductor manufacturing applications. This can be modeled by introducing the mass term into the Hamiltonian which brings new physics related to valley pseudospin. As far as we know, there is very limited work on the subject of atomic collapse difference in $K$ and $K'$ valleys. Ref.~\cite{ref40} reported that a magnetic field modifies the collapse states differently in the two valleys. However, it is not clear whether the atomic collapse in the two valleys is exactly the same. In this work, we will use the tight-binding method to shed light on the atomic collapse in $K$ and $K'$ valleys in gapped graphene. 

The paper is organized as follows. In section II, we present the model and the method used to obtain the relevant quantities. Then the atomic collapse resonances in $K$ valley are studied in section III, and their spatial LDOS and sublattice components are shown in section IV. Next in section V, we study the atomic collapse in $K'$ valley and compare the atomic collapse resonances in the two valleys. Finally, we summarize this study in section VI.     

\section{Model and Method}\label{sec:2}
The tight-binding Hamiltonian of graphene in the presence of a Coulomb center of strength $Z$ is

\begin{multline}
    \hat{H} = \sum_{\langle i,j\rangle}(t_{ij}  \hat{a}_i^\dagger \hat{b}_j + H.c.) + \xi M \sum_i (\hat{a}_i^\dagger \hat{a}-\hat{b}_i^\dagger \hat{b}) 
    \\ -\frac{Ze^2}{\varepsilon} (\sum_i \frac{\hat{a}_i^\dagger \hat{a}_i} {r_i^A}+ \sum_j \frac{\hat{b}_j^\dagger \hat{b}_j} {r_i^B})    
\end{multline}
where $t_{ij}=2.7$ eV is the hopping energy between the nearest neighboring atoms. The operators $\hat{a}_i (a)$ and $\hat{b}_i (b)$ denote creation (annihilation) of an electron on sublattice A and sublattice B, respectively. The second sum is a mass term arising from an on-site energy difference between sublattice A and B. This mass term opens a gap in the band structure and the size of the gap is $\Delta=2M$. Experimentally and theoretically, it has been proven that boron nitride or silicon carbide substrates can open gaps in graphene by breaking sublattice symmetry~\cite{ref41,ref42}. The size of the gap ranges from a few to hundreds of meV. $\xi$ is the valley index. $\xi=1$ indicates the $K$ valley and $\xi=-1$ indicates the $K'$ valleys. The last term is the Coulomb potential. $Z$ is the impurity charge, $\varepsilon$ is the effective dielectric constant and $r_i^{A(B)}$ is the distance from the Coulomb center to the $i$th A(B) sublattice.

In the low-energy excitation regime, the above tight-binding Hamiltonian reduces to the Dirac equation
\begin{equation}
    (-i\hbar v_f \boldsymbol{\sigma} \cdot \boldsymbol{\bigtriangledown} - \frac{Ze^2}{r} + \xi m\sigma_z)\Psi(r)=E\Psi(r)
\end{equation}
where $\hbar$ is the Planck constant and $v_f \approx 10^6m/s$ is the Fermi velocity. $\boldsymbol{\sigma}=(\sigma_x, \sigma_y)$ and $\sigma_z$ are the Pauli matrices. $m$ is the mass of the electron. A non-zero mass brings to the Coulomb problem the appearance of bound states in the gap with energies~\cite{ref38,ref39,ref43}
\begin{equation}
    E_{n,j}=\frac{m}{\hbar v_f}sign(\beta)\frac{n_r+\sqrt{j^2-\beta^2}}{\sqrt{\beta^2+(n_r+\sqrt{j^2-\beta^2})^2}}
\end{equation}
where $\beta=Z\alpha$ is the effective charge which represents the product between the fine structure constant ($\alpha=\frac{e^2}{\hbar v_f}$) of graphene and the value of the charge ($Z$), $n_r=0,1,2,\cdots$ is the principal quantum number and $j$ is the quantum number associated with the total angular momentum quantum operator$J_z=L_x+\frac{1}{2}\sigma_z$. $j$ takes the values $j=\pm \frac{1}{2},\frac{3}{2},\cdots$. The value of the gap is $\Delta =2\frac{m}{\hbar v_f}$.

\begin{figure}
\includegraphics[scale=0.6]{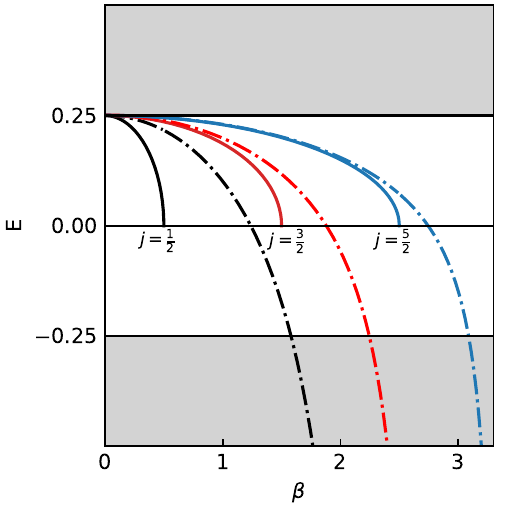}
\caption{\label{fig:fig1} The bound state energy (Eq. (3)) with increasing charge as obtained from the continuum equation. The solid lines represents the evolution of $E_{1j}$ ($j=\frac{1}{2}, \frac{3}{2}$ and $\frac{5}{2}$) for a point nucleus and the solid-dotted lines are for a regularized potential (Adapted from Ref.~\cite{ref43}.)}
\end{figure}

The bound state energies as function of charge are plotted in Fig.~\ref{fig:fig1}. The gray regions represent the positive and negative energy continuum region. The white region is the energy gap. The solid lines represent the evolution of $E_{1j}$ for a point charge. As the charge increases, they are cut off at $E=0$ at $\beta=\frac{1}{2},\frac{3}{2}$ and $\frac{5}{2}$ because when $\beta  > j$, $\sqrt{j^2-\beta^2}$ becomes imaginary. This imaginary energy reflects that the found state turns into a resonant state. In order to solve the above problem, a regularize Coulomb potential is considered at small $r$. When $r<r_0$, the potential is taken constant where {$r_0$ is a regularization parameter as done is Ref.~\cite{ref11}. The dash-dotted lines are for a regularized potential adapted from~\cite{ref43}. 

In order to investigate valley and lattice effects which were neglected previously, we go beyond the continuum approximation and include the discrete lattice structure by using the tight-binding (TB) model. The latter is solved using the open source tight-binding package Pybinding ~\cite{ref44}. In all calculations a hexagonal flake with 200 nm sides is used and an energy broadening of 10 meV is considered. The impurity is put above the graphene plane which removes the singularity of the pure Coulomb potential. For our numerical calculation we take the distance from the impurity to the graphene plane 0.5 $nm$ and the band gap $\Delta=0.5$ eV.

\section{Atomic collapse in $K$ valley}{\label{sec:3}}
\begin{figure}
\includegraphics[scale=0.5]{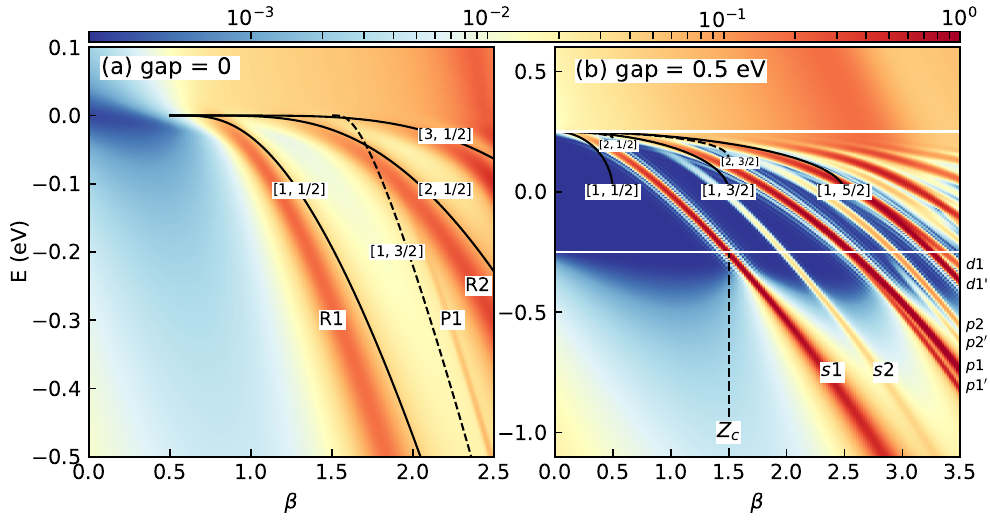}
\caption{\label{fig:fig2} Colormap of the LDOS (in log scale) as a function of charge and energy. (a) LDOS of perfect graphene. The black lines are $E_{nj} \approx \hbar v_f \frac{\beta}{r_0}e^{-\pi n / \sqrt{\beta^2-j^2}}$. (b) LDOS of gapped graphene. The black curves are plotted according to Eq. (3) and labeled with its quantum number. The two white horizontal lines mark the edges of the band gap. The black dashed vertical line marks the critical charge $Z_c$.}
\end{figure}

\begin{figure}
\includegraphics[scale=0.6]{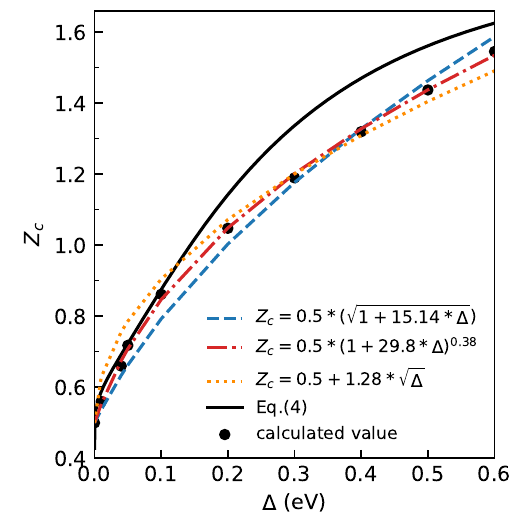}
\caption{\label{fig:fig3} The relation between the supercritical charge $Z_c$ and band gap $\Delta$. $Z_c$ is fitted to three different functions, blue dashed: $Z_c=0.5*\sqrt{1+15.14*\Delta}$, red dashdot: $Z_c=0.5*(1+29.8*\Delta)^{0.38}$, orange dotted: $Z_c=0.5 
+1.28*\sqrt{\Delta}$. The black solid line is the theoretical value from Eq.(4).}
\end{figure}

\begin{figure}
\includegraphics[scale=0.6]{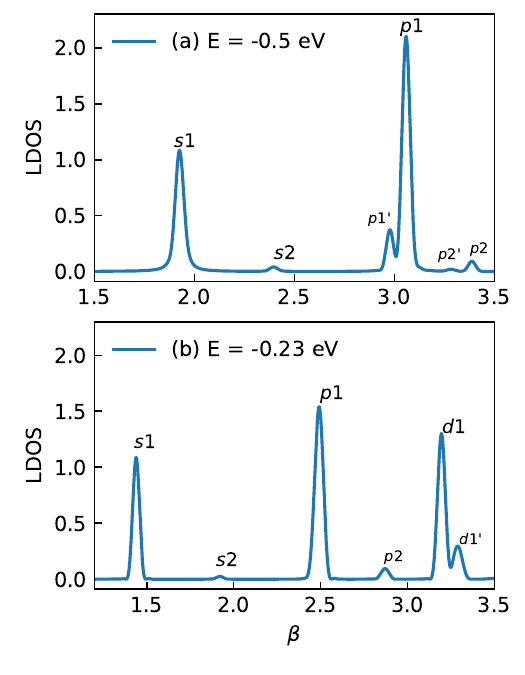}
\caption{\label{fig:fig4} Cut of E = -0.5 eV and E = -0.23 eV in Fig.~\ref{fig:fig2}(b). (a) E = -0.5 eV and (b) E = -0.23 eV.}
\end{figure}

The LDOS at the center of graphene flake, which is just below the Coulomb charge, is plotted as function of energy and charge in Fig.~\ref{fig:fig2}. Fig.~\ref{fig:fig2}(a) is the LDOS of perfect graphene without gap. As the charge increases beyond the supercritical value, the energies of supercritical states behave approximately as $E_{nj} \approx \hbar v_f \frac{\beta}{r_0}e^{-\pi n / \sqrt{\beta^2-j^2}}$~\cite{ref8}. The three lowest states with angular moment quantum number $j=\frac{1}{2}$ are plotted as black solid lines and the dashed line is the lowest state with angular moment quantum number $j=\frac{3}{2}$. The atomic collapse states are recognized as the high LDOS resonances in the negative energy region. Each atomic collapse resonance corresponds to a supercritical state with a specific set of quantum numbers $[n, j]$. The naming of these LDOS resonances are based on the spatial symmetry of their spatial LDOS and are the same as in Ref.~\cite{ref11}. R1 is the 1$s$ atomic collapse with quantum number $[n, j] = [1, \frac{1}{2}]$, R2 is the 2$s$ state with $[n, j] = [2, \frac{1}{2}]$ and P1 is the 1$p$ state with $[n, j] = [2, \frac{3}{2}]$.

$\xi$ is the valley index and we chose $\xi=1$ to plot Fig.~\ref{fig:fig2}(b). Thus Fig.~\ref{fig:fig2}(b) shows the LDOS in $K$ valley of gapped graphene. We chose $M$ = 0.25 eV and the value of the gap is 0.5 eV. The two white horizontal lines mark the upper and lower boundaries of the band gap.  With increaseing of charge, bound states in the gap start to peel off from the bottom of the conductance band, and then fall toward the valence band and after crossing the top of the valence band, drop into the negative energy continuum region and change to atomic collapse quasi-bound states. In this work, the supercritical charge $Z_c$ is defined as the charge at which the lowest bound state crosses the top of the valence band ~\cite{ref38, ref39}. $Z_c$ is highlighted by the vertical dashed line in Fig.~\ref{fig:fig2}(b). 

The numerical obtained values of the supercritical charge $Z_c$ is shown by the solid black dots in Fig.~\ref{fig:fig3} as function of the band gap $\Delta$. At the edge of the boundary of the lower energy continuum $E=-\Delta/2$, the Dirac equation can be solved analytically from which we find the supercritical charge $Z_c$ for the ground state with $n=0, j=\frac{1}{2}$ by solving ~\cite{ref45}
\begin{equation}
    \frac{J_1(Z_c)}{J_0(Z_c)}=\frac{1}{2Z_c}(1-z\frac{K_{iv}^{'}(z)}{K_{iv}(z)})
\end{equation}
where $z=\sqrt{4\Delta r_0 Z_c}$ and $v=\sqrt{4Z_c^2-1}$. $J_l(x)$ is the Bessel function of the first kind and $K_{iv}(x)$ is the modified Bessel function of imaginary order. The black solid line in Fig.~\ref{fig:fig3} is the solution of Eq. (4) and agree very well with our TB results for $\Delta <0.5$ eV. The black dots are calculated values and fitted to three functions: $Z_c=0.5*\sqrt{1+15.14*\Delta}$, $Z_c=0.5+1.28*\sqrt{\Delta}$ and $Z_c=0.5*(1+29.8*\Delta)^{0.38}$. The latter gives the best fit.

Compared with Fig.~\ref{fig:fig2}(a), another obvious feature of Fig.~\ref{fig:fig2}(b) is that the LDOS resonances are divided into two groups: strong with higher intensity and weak with lower intensity. In order to find out the differences between these two sets of LDOS resonances and clarify the relation between the LDOS resonances and the bound states, we added in Fig.~\ref{fig:fig2}(b) the bound states in the band gap using Eq. (3). A set of quantum numbers $[n,j]$ is used to label the bound state. The bound states with principal quantum numbers $n=1$ are plotted in black solid lines and the bound states with principal quantum number $n=2$ are plotted in black dashed lines. These bound states are truncated at $\beta=j$ and the reason has been explained in section II. From this it appears clearly that the LDOS resonances with high intensity are resulting from the bound states $n=1$ while the weak collapse resonances are from the bound states with $n=2$. 

We gave each LDOS resonance a label according to the quantum numbers of its corresponding bound state. $s$1 corresponds to the state $[n,j]$ = $[1, \frac{1}{2}]$ and $s$2 corresponds to the state $[n,j]$ = $[2, \frac{1}{2}]$. $p$1 and $p1'$ correspond to the state $[n,j]$ = $[1, \frac{3}{2}]$, and $p$2 and $p2'$ correspond to the state $[n,j]$ = $[2, \frac{3}{2}]$. $d1$ and $d1'$ are higher order states with larger $j$, i. e. $[n,j]$ = $[1, \frac{5}{2}]$. The $p1'$ and $p2'$ resonances appear only in the continuum regime as a splitting of respectively, $p1$ and $p2$. $d1'$ resonance is the splitting of $d1$ state and notice that the splitting of $d$ state begins already within the band gap. These splittings of the resonances are shown clearly in Fig.~\ref{fig:fig4} by the cuts at E = -0.5 eV and -0.23 eV of Fig.~\ref{fig:fig2}(b). Notice that the $s$ state resonances with quantum number $j$ = $\frac{1}{2}$ do not split. While higher order resonances show a splitting as the charge increases and the larger the quantum number $j$ the earlier the splitting begins.  

Next, we study the spatial LDOS of these collapse resonances to clarify the relation between collapse resonance states and atomic orbitals. 

\section{The spatial LDOS of different atomic collapse resonances}\label{sec:4}

The spatial LDOS of $s$1 and $s$2 collapse resonances are shown in Fig.~\ref{fig:fig5}. The left column shows the total spatial LDOS, and the middle right columns show the sublattice components to the spatial LDOS. For the $s$1 resonance in Fig.~\ref{fig:fig5}(a1), its spatial LDOS is highest at the center and decreases along the radial direction typical for a $s$1 hydrogen atom state of a Coulomb potential. While the spatial LDOS of $s$2 resonance has in addition a node which is analogous to the atomic 2$s$ state. Because the onsite energies of the two sublattices are equal but opposite in sign, the sublattice symmetry is broken. Thus the two sublattices will contribute differently to the spatial LDOS. We show the sublattice components separately in Figs.~\ref{fig:fig5}(a2, a3, b2 and b3). For the 1$s$ atomic collapse state, its spatial LDOS of A-sublattice component in Fig.~\ref{fig:fig5}(a2) and B-sublattice component in Fig.~\ref{fig:fig5}(a3) look like the one from a Coulomb potential, but the contribution of B-sublattice is smaller than that from A-sublattice. For the 2$s$ atomic collapse state, its spatial LDOS distribution on both sublattices in Figs.~\ref{fig:fig5}(b2 and b3) look also like the one of a Coulomb potential but the lower intensity node which is the symbolic feature of 2$s$ atomic orbital is only observed on B-sublattice in Fig.~\ref{fig:fig5}(b3). Notice that the LDOS is not perfectly circular symmetry and contains signatures of the discrete graphene.

\begin{figure}
\includegraphics[scale=0.5]{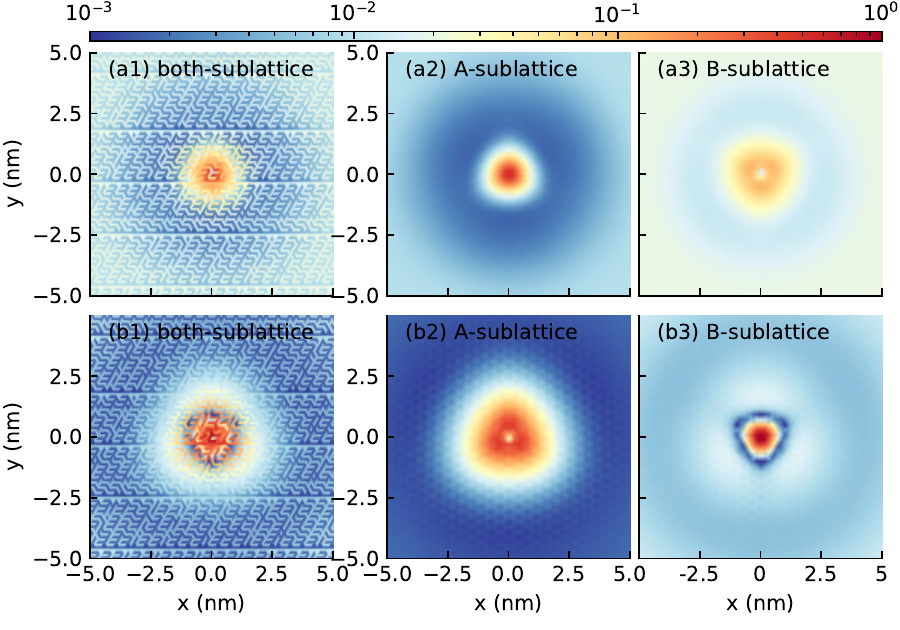}
\caption{\label{fig:fig5} Spatial LDOS (in log scale) of $s$1 and $s$2 collapse resonances. (a1-a3) $(E,\beta)$=(-0.85 eV, 2.5) for $s$1 collapse resonances and (b1-b3) $(E,\beta)$=(-0.55 eV, 2.5) for $s$2 collapse resonances. Left column shows the total LDOS and the sublattice components are plotted in the middle (A-sublattice) and the right (B-sublattice) columns. }
\end{figure}

\begin{figure}
\includegraphics[scale=0.5]{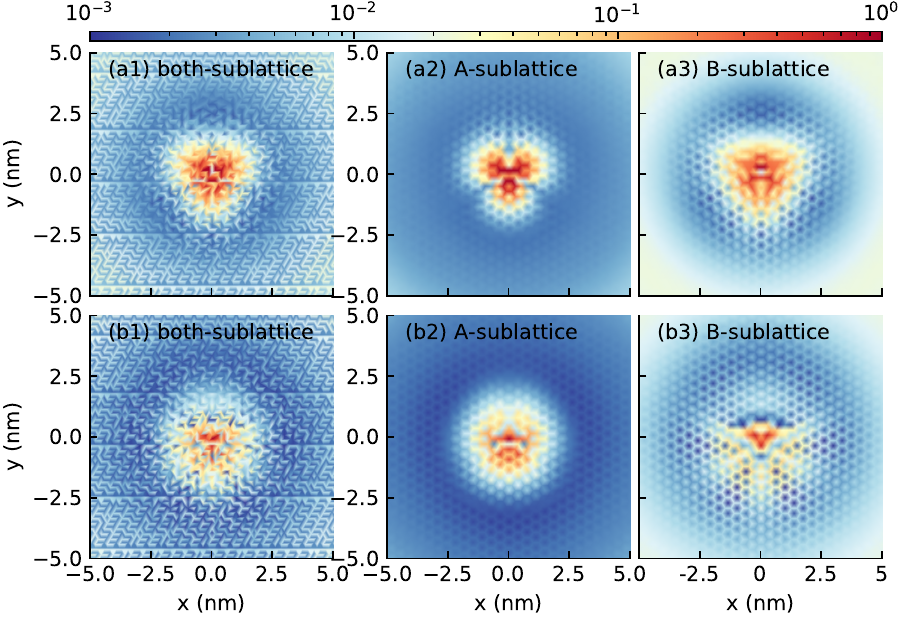}
\caption{\label{fig:fig6} Spatial LDOS (in log scale) of $p$1 and $p1'$ collapse resonances. (a1-a3) $(E,\beta)$=(-0.85 eV, 3.5) for $p1'$ collapse resonance and (b1-b3) $(E,\beta)$=(-0.75 eV, 3.5) for $p1$ collapse resonance. Left column shows the total LDOS and the sublattice components are plotted in the middle (A-sublattice) and the right (B-sublattice) columns.}
\end{figure}

\begin{figure}
\includegraphics[scale=0.5]{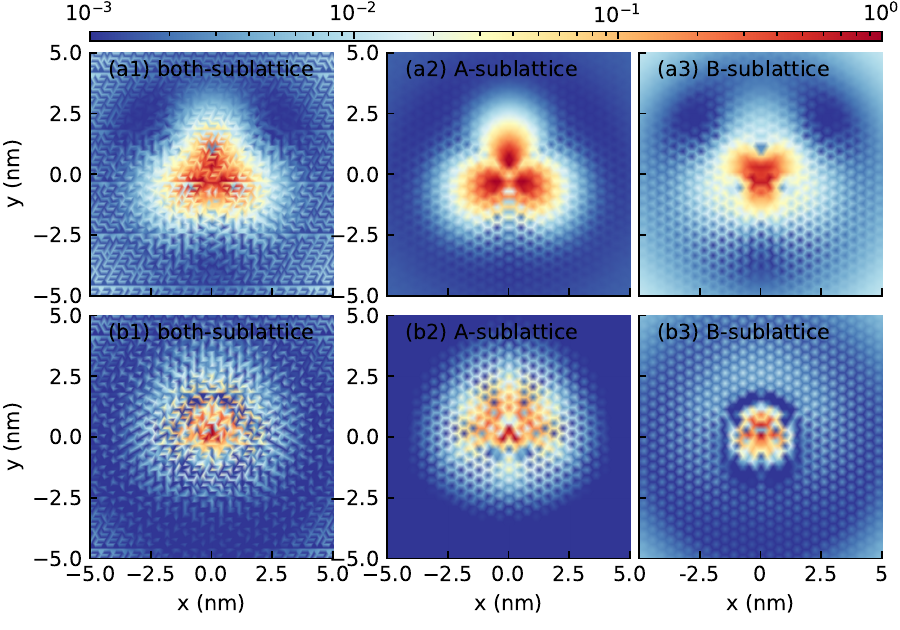}
\caption{\label{fig:fig7} Spatial LDOS (in log scale) of $p$2 and $p2'$ collapse resonances. (a1-a3) $(E,\beta)$=(-0.62 eV, 3.5) for $p2'$ collapse resonance and (b1-b3) $(E,\beta)$=(-0.55 eV, 3.5) for $p2$ collapse resonance. Left column shows the total LDOS and the sublattice components are plotted in the middle (A-sublattice) and the right (B-sublattice) columns. }
\end{figure}

In Fig.~\ref{fig:fig6} we show the spatial LDOS of $p$1 and $p1'$ resonances and the individual sublattice components. Unlike the spatial LDOS of the $s$ orbital collapse resonances in Fig.~\ref{fig:fig5}, the total spatial LDOS in Figs.~\ref{fig:fig6}(a1 and b1) do not look like those of a $p$ state of a single Coulomb potential. By separating the two sublattice components from the total spatial LDOS, this feature is more obvious in Figs.~\ref{fig:fig6}(a2, a3, b2 and b3). The A-sublattice component of $p1'$ resonance in Fig.~\ref{fig:fig6}(a2) show a lobe structure. Because graphene is a 2D plane and the A-sublattice is a triangular lattice, the usual $p_x$ and $p_y$ orbitals are not found separately but linear combination appears such that it exhibits 3-fold symmetry. Different from the A-sublattice component, the spatial LDOS of $p1'$ resonance on B-sublattice in Fig.~\ref{fig:fig6}(a3) show a more complicated symmetry. For the $p1$ resonance, its A-sublattice component in Fig.~\ref{fig:fig6}(b2) does not show the characteristics of either a $p$ or $s$ atomic orbital while its B-sublattice component in Fig.~\ref{fig:fig6}(b3) show a lobe structure and is symmetric about $x=0$. 

The spatial LDOS of a higher order $p$ atomic collapse resonance is shown in Fig.~\ref{fig:fig7}. The above conclusions of $p$ atomic orbital are proven to be true again. We take a look at the sublattice component first. The spatial LDOS on A-sublattice of $p2'$ resonance in Fig.~\ref{fig:fig7}(a2) has the same characteristics as the $p1'$ resonance in Fig.~\ref{fig:fig6}(a2) except that the spatial LDOS image is flipped upside down. Meanwhile the spatial LDOS on B-sublattice of $p2'$ resonance in Fig.~\ref{fig:fig7}(a3) also show 3-fold symmetry, but in addition it has three lower LDOS intensity nodes. For the $p2$ resonance in Figs. ~\ref{fig:fig7}(b2 and b3), the lobe structure is observed on A-sublattice component and the lower LDOS intensity node is observed on B-sublattice. These features together prove that $p$2 and $p2'$ states are higher-order $p$ atomic orbitals compared to the $p1$ and $p1'$ states. The total spatial LDOS in Figs.~\ref{fig:fig7}(a1 and b1) show the combined features of A-sublattice component and B-sublattice component. 

Usually, $p$ atomic orbital is expressed in terms of $p_x$, $p_y$ and $p_z$ orbitals, and shows symmetry in $x$, $y$ and $z$ axes. Since graphene is a 2D plane, it is reasonable that the counterpart of $p$ atomic orbital in graphene can be expressed in terms of only $p$ and $p'$ orbitals. Our analysis of $p$ and $p'$ collapse resonances in Figs.~\ref{fig:fig6} and ~\ref{fig:fig7} demonstrate that $p$ and $p'$ orbital exhibit in-plane symmetry in different directions. Notice that $s$2 and $p1$ bound states are no longer degenerate in energy, as for a non-relativistic Coulomb problem, which is due to the discrete lattice structure and the fact that we have here a relativistic-type of Coulomb problem. Now it is essential to point out that more atomic collapse states are observed than the bound states found in the gap. The extra atomic collapse states stem from the states which are the $p$ or $p'$ components of a $p$ atomic orbital with $j=
\frac{3}{2}$. In Fig.~\ref{fig:fig2}(b), it shows very clearly that $p1$ and $p1'$ atomic collapse resonances are transformed from the same bound state with $[n,j]=[1,\frac{3}{2}]$. Similar conclusions hold for the $p2$ atomic collapse resonance state. 

\begin{figure}
\includegraphics[scale=0.7]{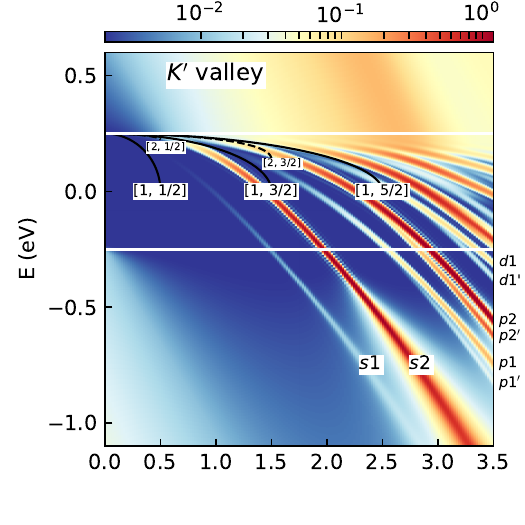}
\caption{\label{fig:fig8} Colormap of the LDOS (in logscale) in $K'$ valley as a function of energy and charge. $\xi=-1$. The other parameters are the same as in Fig.~\ref{fig:fig2}(b).}
\end{figure}

\begin{figure}
\includegraphics[scale=0.7]{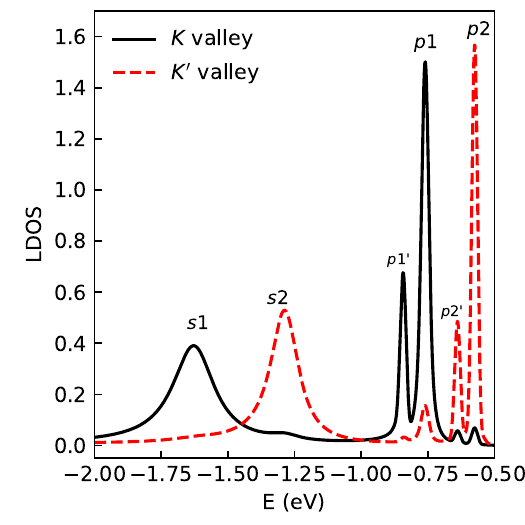}
\caption{\label{fig:fig9} Cut of $\beta=3.5$ in Figs.~\ref{fig:fig2}(b) and Fig.~\ref{fig:fig8}. The black solid line is the LDOS in $K$ valley from Fig.~\ref{fig:fig2}(b). The red dashed line is the LDOS in $K'$ valley from Fig.~\ref{fig:fig8}.}
\end{figure}

\section{The atomic collapse in $K'$ valley}\label{sec:5}

In all the above discussion, we only considered the atomic collapse in $K$ valley with $\xi=1$. In Fig.~\ref{fig:fig7}, the atomic collapse resonance in $K'$ valley is investigated by setting $\xi=-1$ and the other parameters are kept the same as for Fig.~\ref{fig:fig2}(b). The LDOS resonances in $K'$ valley are almost the same as the LDOS resonances in $K$ valley.  

By making a cut along the charge at $\beta=3.5$ in Fig.~\ref{fig:fig2}(b) and Fig.~\ref{fig:fig8}, the LDOS as a function of energy is plotted in Fig.~\ref{fig:fig9}. The LDOS resonance peaks in $K$ valley and $K'$ valley are found at the same energies. But the strong (weak) LDOS resonances in $K$ valley change to weak (strong) LDOS resonances in $K'$ valley. Valley index is an intrinsic degree of freedom, which can be used to store and transfer information as electron spin, which are here described as pseudospin. The LDOS resonances in the two valleys show opposite strong and weak characteristics which has not been pointed out previously. This is a pseudospin polarize phenomenon which has potential applications in valleytronics.

\section{Conclusion}\label{sec:6} 
In this work, we investigated the atomic collapse resonances in gapped graphene. The gap in graphene can e. g. be opened by substrate engineering. A series of bound states with different quantum numbers are found in the gap. With increase of charge of the impurity, these bound states pass through the gap, fall into the negative continuum energy region and turn into atomic collapse resonances. The supercritical charge $Z_c$ increases sub-lineary with the band gap $\Delta$ and could be best fitted to $Z_c=0.5*(1+29.8*\Delta (eV))^{0.38}$. A close correspondence between the collapse resonances and atomic orbitals is established. Different from previous studies of atomic collapse in graphene, the collapse state of the $p$ atomic orbital split into $p$ and $p'$ LDOS resonances in the supercritical region. The $p$ and $p'$ orbitals can be regarded as the two components of $p$ atomic orbital in the graphene plane whose degeneracy is now lifted. By a spatial LDOS analysis, we find that the LDOS in the A and B sublattices exhibit in-plane symmetry along different directions. Finally, the most important conclusion of this work is that the atomic collapse phenomenons in the two valleys are different. A strong(weak) LDOS resonance in $K$ valley is a weak(strong) LDOS resonance in $K'$ valley. This phenomenon can be regarded as a kind of valley polarization and has potential application in valleytronics.

\begin{acknowledgments}
This work was supported by the National Natural Science Foundation of China (Grant No. 62004053, U24A20296)).
\end{acknowledgments} 

\bibliography{references}

\begin{thebibliography}{45}%
\makeatletter
\providecommand \@ifxundefined [1]{%
 \@ifx{#1\undefined}
}%
\providecommand \@ifnum [1]{%
 \ifnum #1\expandafter \@firstoftwo
 \else \expandafter \@secondoftwo
 \fi
}%
\providecommand \@ifx [1]{%
 \ifx #1\expandafter \@firstoftwo
 \else \expandafter \@secondoftwo
 \fi
}%
\providecommand \natexlab [1]{#1}%
\providecommand \enquote  [1]{``#1''}%
\providecommand \bibnamefont  [1]{#1}%
\providecommand \bibfnamefont [1]{#1}%
\providecommand \citenamefont [1]{#1}%
\providecommand \href@noop [0]{\@secondoftwo}%
\providecommand \href [0]{\begingroup \@sanitize@url \@href}%
\providecommand \@href[1]{\@@startlink{#1}\@@href}%
\providecommand \@@href[1]{\endgroup#1\@@endlink}%
\providecommand \@sanitize@url [0]{\catcode `\\12\catcode `\$12\catcode `\&12\catcode `\#12\catcode `\^12\catcode `\_12\catcode `\%12\relax}%
\providecommand \@@startlink[1]{}%
\providecommand \@@endlink[0]{}%
\providecommand \url  [0]{\begingroup\@sanitize@url \@url }%
\providecommand \@url [1]{\endgroup\@href {#1}{\urlprefix }}%
\providecommand \urlprefix  [0]{URL }%
\providecommand \Eprint [0]{\href }%
\providecommand \doibase [0]{https://doi.org/}%
\providecommand \selectlanguage [0]{\@gobble}%
\providecommand \bibinfo  [0]{\@secondoftwo}%
\providecommand \bibfield  [0]{\@secondoftwo}%
\providecommand \translation [1]{[#1]}%
\providecommand \BibitemOpen [0]{}%
\providecommand \bibitemStop [0]{}%
\providecommand \bibitemNoStop [0]{.\EOS\space}%
\providecommand \EOS [0]{\spacefactor3000\relax}%
\providecommand \BibitemShut  [1]{\csname bibitem#1\endcsname}%
\let\auto@bib@innerbib\@empty
\bibitem [{\citenamefont {Novoselov}\ \emph {et~al.}(2004)\citenamefont {Novoselov}, \citenamefont {Geim}, \citenamefont {Morozov}, \citenamefont {Jiang}, \citenamefont {Zhang}, \citenamefont {Dubonos}, \citenamefont {Grigorieva},\ and\ \citenamefont {A.}}]{ref1}%
  \BibitemOpen
  \bibfield  {author} {\bibinfo {author} {\bibfnamefont {K.~S.}\ \bibnamefont {Novoselov}}, \bibinfo {author} {\bibfnamefont {A.~K.}\ \bibnamefont {Geim}}, \bibinfo {author} {\bibfnamefont {S.~V.}\ \bibnamefont {Morozov}}, \bibinfo {author} {\bibfnamefont {D.}~\bibnamefont {Jiang}}, \bibinfo {author} {\bibfnamefont {Y.}~\bibnamefont {Zhang}}, \bibinfo {author} {\bibfnamefont {S.~V.}\ \bibnamefont {Dubonos}}, \bibinfo {author} {\bibfnamefont {I.~V.}\ \bibnamefont {Grigorieva}},\ and\ \bibinfo {author} {\bibfnamefont {F.~A.}\ \bibnamefont {A.}},\ }\href {https://doi.org/10.1126/science.1102896} {\bibfield  {journal} {\bibinfo  {journal} {Science}\ }\textbf {\bibinfo {volume} {306}},\ \bibinfo {pages} {666} (\bibinfo {year} {2004})}\BibitemShut {NoStop}%
\bibitem [{\citenamefont {Castro~Neto}\ \emph {et~al.}(2009)\citenamefont {Castro~Neto}, \citenamefont {Guinea}, \citenamefont {Peres}, \citenamefont {Novoselov},\ and\ \citenamefont {Geim}}]{ref2}%
  \BibitemOpen
  \bibfield  {author} {\bibinfo {author} {\bibfnamefont {A.~H.}\ \bibnamefont {Castro~Neto}}, \bibinfo {author} {\bibfnamefont {F.}~\bibnamefont {Guinea}}, \bibinfo {author} {\bibfnamefont {N.~M.~R.}\ \bibnamefont {Peres}}, \bibinfo {author} {\bibfnamefont {K.~S.}\ \bibnamefont {Novoselov}},\ and\ \bibinfo {author} {\bibfnamefont {A.~K.}\ \bibnamefont {Geim}},\ }\href {https://doi.org/10.1103/RevModPhys.81.109} {\bibfield  {journal} {\bibinfo  {journal} {Rev. Mod. Phys.}\ }\textbf {\bibinfo {volume} {81}},\ \bibinfo {pages} {109} (\bibinfo {year} {2009})}\BibitemShut {NoStop}%
\bibitem [{\citenamefont {Reinhardt}\ and\ \citenamefont {Greiner}(1997{\natexlab{a}})}]{ref3}%
  \BibitemOpen
  \bibfield  {author} {\bibinfo {author} {\bibfnamefont {J.}~\bibnamefont {Reinhardt}}\ and\ \bibinfo {author} {\bibfnamefont {W.}~\bibnamefont {Greiner}},\ }\href {https://doi.org/10.1088/0034-4885/40/3/001} {\bibfield  {journal} {\bibinfo  {journal} {Rep. Prog. Phys.}\ }\textbf {\bibinfo {volume} {40}},\ \bibinfo {pages} {219} (\bibinfo {year} {1997}{\natexlab{a}})}\BibitemShut {NoStop}%
\bibitem [{\citenamefont {Reinhardt}\ and\ \citenamefont {Greiner}(1997{\natexlab{b}})}]{ref4}%
  \BibitemOpen
  \bibfield  {author} {\bibinfo {author} {\bibfnamefont {J.}~\bibnamefont {Reinhardt}}\ and\ \bibinfo {author} {\bibfnamefont {W.}~\bibnamefont {Greiner}},\ }\href {https://doi.org/10.1088/0034-4885/40/3/001} {\bibfield  {journal} {\bibinfo  {journal} {Rep. Prog. Phys.}\ }\textbf {\bibinfo {volume} {40}},\ \bibinfo {pages} {219} (\bibinfo {year} {1997}{\natexlab{b}})}\BibitemShut {NoStop}%
\bibitem [{\citenamefont {Schweppe}\ \emph {et~al.}(1983)\citenamefont {Schweppe}, \citenamefont {Gruppe}, \citenamefont {Bethge}, \citenamefont {Bokemeyer}, \citenamefont {Cowan}, \citenamefont {Folger}, \citenamefont {Greenberg}, \citenamefont {Grein}, \citenamefont {Ito}, \citenamefont {Schule}, \citenamefont {Schwalm}, \citenamefont {Stiebing}, \citenamefont {Trautmann}, \citenamefont {Vincent},\ and\ \citenamefont {Waldschmidt}}]{ref5}%
  \BibitemOpen
  \bibfield  {author} {\bibinfo {author} {\bibfnamefont {J.}~\bibnamefont {Schweppe}}, \bibinfo {author} {\bibfnamefont {A.}~\bibnamefont {Gruppe}}, \bibinfo {author} {\bibfnamefont {K.}~\bibnamefont {Bethge}}, \bibinfo {author} {\bibfnamefont {H.}~\bibnamefont {Bokemeyer}}, \bibinfo {author} {\bibfnamefont {T.}~\bibnamefont {Cowan}}, \bibinfo {author} {\bibfnamefont {H.}~\bibnamefont {Folger}}, \bibinfo {author} {\bibfnamefont {J.~S.}\ \bibnamefont {Greenberg}}, \bibinfo {author} {\bibfnamefont {H.}~\bibnamefont {Grein}}, \bibinfo {author} {\bibfnamefont {S.}~\bibnamefont {Ito}}, \bibinfo {author} {\bibfnamefont {R.}~\bibnamefont {Schule}}, \bibinfo {author} {\bibfnamefont {D.}~\bibnamefont {Schwalm}}, \bibinfo {author} {\bibfnamefont {K.~E.}\ \bibnamefont {Stiebing}}, \bibinfo {author} {\bibfnamefont {N.}~\bibnamefont {Trautmann}}, \bibinfo {author} {\bibfnamefont {P.}~\bibnamefont {Vincent}},\ and\ \bibinfo {author} {\bibfnamefont {M.}~\bibnamefont {Waldschmidt}},\ }\href
  {https://doi.org/10.1103/PhysRevLett.51.2261} {\bibfield  {journal} {\bibinfo  {journal} {Phys. Rev. Lett.}\ }\textbf {\bibinfo {volume} {51}},\ \bibinfo {pages} {2261} (\bibinfo {year} {1983})}\BibitemShut {NoStop}%
\bibitem [{\citenamefont {Cowan}\ \emph {et~al.}(1985)\citenamefont {Cowan}, \citenamefont {Backe}, \citenamefont {Begemann}, \citenamefont {Bethge}, \citenamefont {Bokemeyer}, \citenamefont {Folger}, \citenamefont {Greenberg}, \citenamefont {Grein}, \citenamefont {Gruppe}, \citenamefont {Kido}, \citenamefont {Kl\"uver}, \citenamefont {Schwalm}, \citenamefont {Schweppe}, \citenamefont {Stiebing}, \citenamefont {Trautmann},\ and\ \citenamefont {Vincent}}]{ref6}%
  \BibitemOpen
  \bibfield  {author} {\bibinfo {author} {\bibfnamefont {T.}~\bibnamefont {Cowan}}, \bibinfo {author} {\bibfnamefont {H.}~\bibnamefont {Backe}}, \bibinfo {author} {\bibfnamefont {M.}~\bibnamefont {Begemann}}, \bibinfo {author} {\bibfnamefont {K.}~\bibnamefont {Bethge}}, \bibinfo {author} {\bibfnamefont {H.}~\bibnamefont {Bokemeyer}}, \bibinfo {author} {\bibfnamefont {H.}~\bibnamefont {Folger}}, \bibinfo {author} {\bibfnamefont {J.~S.}\ \bibnamefont {Greenberg}}, \bibinfo {author} {\bibfnamefont {H.}~\bibnamefont {Grein}}, \bibinfo {author} {\bibfnamefont {A.}~\bibnamefont {Gruppe}}, \bibinfo {author} {\bibfnamefont {Y.}~\bibnamefont {Kido}}, \bibinfo {author} {\bibfnamefont {M.}~\bibnamefont {Kl\"uver}}, \bibinfo {author} {\bibfnamefont {D.}~\bibnamefont {Schwalm}}, \bibinfo {author} {\bibfnamefont {J.}~\bibnamefont {Schweppe}}, \bibinfo {author} {\bibfnamefont {K.~E.}\ \bibnamefont {Stiebing}}, \bibinfo {author} {\bibfnamefont {N.}~\bibnamefont {Trautmann}},\ and\ \bibinfo {author} {\bibfnamefont
  {P.}~\bibnamefont {Vincent}},\ }\href {https://doi.org/10.1103/PhysRevLett.54.1761} {\bibfield  {journal} {\bibinfo  {journal} {Phys. Rev. Lett.}\ }\textbf {\bibinfo {volume} {54}},\ \bibinfo {pages} {1761} (\bibinfo {year} {1985})}\BibitemShut {NoStop}%
\bibitem [{\citenamefont {Zhang}\ and\ \citenamefont {Fogler}(2008)}]{ref7}%
  \BibitemOpen
  \bibfield  {author} {\bibinfo {author} {\bibfnamefont {L.~M.}\ \bibnamefont {Zhang}}\ and\ \bibinfo {author} {\bibfnamefont {M.~M.}\ \bibnamefont {Fogler}},\ }\href {https://doi.org/10.1103/PhysRevLett.100.116804} {\bibfield  {journal} {\bibinfo  {journal} {Phys. Rev. Lett.}\ }\textbf {\bibinfo {volume} {100}},\ \bibinfo {pages} {116804} (\bibinfo {year} {2008})}\BibitemShut {NoStop}%
\bibitem [{\citenamefont {Shytov}\ \emph {et~al.}(2007{\natexlab{a}})\citenamefont {Shytov}, \citenamefont {Katsnelson},\ and\ \citenamefont {Levitov}}]{ref8}%
  \BibitemOpen
  \bibfield  {author} {\bibinfo {author} {\bibfnamefont {A.~V.}\ \bibnamefont {Shytov}}, \bibinfo {author} {\bibfnamefont {M.~I.}\ \bibnamefont {Katsnelson}},\ and\ \bibinfo {author} {\bibfnamefont {L.~S.}\ \bibnamefont {Levitov}},\ }\href {https://doi.org/10.1103/PhysRevLett.99.246802} {\bibfield  {journal} {\bibinfo  {journal} {Phys. Rev. Lett.}\ }\textbf {\bibinfo {volume} {99}},\ \bibinfo {pages} {246802} (\bibinfo {year} {2007}{\natexlab{a}})}\BibitemShut {NoStop}%
\bibitem [{\citenamefont {Shytov}\ \emph {et~al.}(2007{\natexlab{b}})\citenamefont {Shytov}, \citenamefont {Katsnelson},\ and\ \citenamefont {Levitov}}]{ref9}%
  \BibitemOpen
  \bibfield  {author} {\bibinfo {author} {\bibfnamefont {A.~V.}\ \bibnamefont {Shytov}}, \bibinfo {author} {\bibfnamefont {M.~I.}\ \bibnamefont {Katsnelson}},\ and\ \bibinfo {author} {\bibfnamefont {L.~S.}\ \bibnamefont {Levitov}},\ }\href {https://doi.org/10.1103/PhysRevLett.99.236801} {\bibfield  {journal} {\bibinfo  {journal} {Phys. Rev. Lett.}\ }\textbf {\bibinfo {volume} {99}},\ \bibinfo {pages} {236801} (\bibinfo {year} {2007}{\natexlab{b}})}\BibitemShut {NoStop}%
\bibitem [{\citenamefont {Wang}\ \emph {et~al.}(2013)\citenamefont {Wang}, \citenamefont {Wong}, \citenamefont {Shytov}, \citenamefont {Brar}, \citenamefont {Choi}, \citenamefont {Wu}, \citenamefont {Tsai}, \citenamefont {Regan}, \citenamefont {Zettl}, \citenamefont {Kawakami}, \citenamefont {Louie}, \citenamefont {Levitov},\ and\ \citenamefont {Crommie}}]{ref10}%
  \BibitemOpen
  \bibfield  {author} {\bibinfo {author} {\bibfnamefont {Y.}~\bibnamefont {Wang}}, \bibinfo {author} {\bibfnamefont {D.}~\bibnamefont {Wong}}, \bibinfo {author} {\bibfnamefont {A.~V.}\ \bibnamefont {Shytov}}, \bibinfo {author} {\bibfnamefont {V.~W.}\ \bibnamefont {Brar}}, \bibinfo {author} {\bibfnamefont {S.}~\bibnamefont {Choi}}, \bibinfo {author} {\bibfnamefont {Q.}~\bibnamefont {Wu}}, \bibinfo {author} {\bibfnamefont {H.-Z.}\ \bibnamefont {Tsai}}, \bibinfo {author} {\bibfnamefont {W.}~\bibnamefont {Regan}}, \bibinfo {author} {\bibfnamefont {A.}~\bibnamefont {Zettl}}, \bibinfo {author} {\bibfnamefont {R.~K.}\ \bibnamefont {Kawakami}}, \bibinfo {author} {\bibfnamefont {S.~G.}\ \bibnamefont {Louie}}, \bibinfo {author} {\bibfnamefont {L.~S.}\ \bibnamefont {Levitov}},\ and\ \bibinfo {author} {\bibfnamefont {M.~F.}\ \bibnamefont {Crommie}},\ }\href {https://doi.org/10.1126/science.1234320} {\bibfield  {journal} {\bibinfo  {journal} {Science}\ }\textbf {\bibinfo {volume} {340}},\ \bibinfo {pages} {734} (\bibinfo
  {year} {2013})}\BibitemShut {NoStop}%
\bibitem [{\citenamefont {Mao}\ \emph {et~al.}(2016)\citenamefont {Mao}, \citenamefont {Jiang}, \citenamefont {Moldovan}, \citenamefont {Li}, \citenamefont {Watanabe}, \citenamefont {Taniguchi}, \citenamefont {Masir}, \citenamefont {Peeters},\ and\ \citenamefont {Andrei}}]{ref11}%
  \BibitemOpen
  \bibfield  {author} {\bibinfo {author} {\bibfnamefont {J.}~\bibnamefont {Mao}}, \bibinfo {author} {\bibfnamefont {Y.}~\bibnamefont {Jiang}}, \bibinfo {author} {\bibfnamefont {D.}~\bibnamefont {Moldovan}}, \bibinfo {author} {\bibfnamefont {G.}~\bibnamefont {Li}}, \bibinfo {author} {\bibfnamefont {K.}~\bibnamefont {Watanabe}}, \bibinfo {author} {\bibfnamefont {T.}~\bibnamefont {Taniguchi}}, \bibinfo {author} {\bibfnamefont {M.~R.}\ \bibnamefont {Masir}}, \bibinfo {author} {\bibfnamefont {F.~M.}\ \bibnamefont {Peeters}},\ and\ \bibinfo {author} {\bibfnamefont {E.~Y.}\ \bibnamefont {Andrei}},\ }\href {https://doi.org/10.1038/nphys3665} {\bibfield  {journal} {\bibinfo  {journal} {Nat. Phys.}\ }\textbf {\bibinfo {volume} {12}},\ \bibinfo {pages} {545} (\bibinfo {year} {2016})}\BibitemShut {NoStop}%
\bibitem [{\citenamefont {Jiang}\ \emph {et~al.}(2017)\citenamefont {Jiang}, \citenamefont {Mao}, \citenamefont {Moldovan}, \citenamefont {Masir}, \citenamefont {Li}, \citenamefont {Watanabe}, \citenamefont {Taniguchi}, \citenamefont {Peeters},\ and\ \citenamefont {Andrei}}]{ref12}%
  \BibitemOpen
  \bibfield  {author} {\bibinfo {author} {\bibfnamefont {Y.}~\bibnamefont {Jiang}}, \bibinfo {author} {\bibfnamefont {J.}~\bibnamefont {Mao}}, \bibinfo {author} {\bibfnamefont {D.}~\bibnamefont {Moldovan}}, \bibinfo {author} {\bibfnamefont {M.~R.}\ \bibnamefont {Masir}}, \bibinfo {author} {\bibfnamefont {G.}~\bibnamefont {Li}}, \bibinfo {author} {\bibfnamefont {K.}~\bibnamefont {Watanabe}}, \bibinfo {author} {\bibfnamefont {T.}~\bibnamefont {Taniguchi}}, \bibinfo {author} {\bibfnamefont {F.~M.}\ \bibnamefont {Peeters}},\ and\ \bibinfo {author} {\bibfnamefont {E.~Y.}\ \bibnamefont {Andrei}},\ }\href {https://doi.org/10.1038/nnano.2017.181} {\bibfield  {journal} {\bibinfo  {journal} {Nat. Nanotechnol.}\ }\textbf {\bibinfo {volume} {12}},\ \bibinfo {pages} {1045} (\bibinfo {year} {2017})}\BibitemShut {NoStop}%
\bibitem [{\citenamefont {Lu}\ \emph {et~al.}(2019{\natexlab{a}})\citenamefont {Lu}, \citenamefont {Tsai}, \citenamefont {Tatan}, \citenamefont {Wickenburg}, \citenamefont {Omrani}, \citenamefont {Wong}, \citenamefont {Riss}, \citenamefont {Piatti}, \citenamefont {Watanabe}, \citenamefont {Taniguchi}, \citenamefont {Zettl}, \citenamefont {Pereira},\ and\ \citenamefont {Crommie}}]{ref13}%
  \BibitemOpen
  \bibfield  {author} {\bibinfo {author} {\bibfnamefont {J.}~\bibnamefont {Lu}}, \bibinfo {author} {\bibfnamefont {H.-Z.}\ \bibnamefont {Tsai}}, \bibinfo {author} {\bibfnamefont {A.~N.}\ \bibnamefont {Tatan}}, \bibinfo {author} {\bibfnamefont {S.}~\bibnamefont {Wickenburg}}, \bibinfo {author} {\bibfnamefont {A.~A.}\ \bibnamefont {Omrani}}, \bibinfo {author} {\bibfnamefont {D.}~\bibnamefont {Wong}}, \bibinfo {author} {\bibfnamefont {A.}~\bibnamefont {Riss}}, \bibinfo {author} {\bibfnamefont {E.}~\bibnamefont {Piatti}}, \bibinfo {author} {\bibfnamefont {K.}~\bibnamefont {Watanabe}}, \bibinfo {author} {\bibfnamefont {T.}~\bibnamefont {Taniguchi}}, \bibinfo {author} {\bibfnamefont {A.}~\bibnamefont {Zettl}}, \bibinfo {author} {\bibfnamefont {V.~M.}\ \bibnamefont {Pereira}},\ and\ \bibinfo {author} {\bibfnamefont {M.~F.}\ \bibnamefont {Crommie}},\ }\href {https://doi.org/10.1038/s41467-019-08371-2} {\bibfield  {journal} {\bibinfo  {journal} {Nat. Commun.}\ }\textbf {\bibinfo {volume} {10}},\ \bibinfo {pages} {477}
  (\bibinfo {year} {2019}{\natexlab{a}})}\BibitemShut {NoStop}%
\bibitem [{\citenamefont {Pottelberge}\ \emph {et~al.}(2019)\citenamefont {Pottelberge}, \citenamefont {Moldovan}, \citenamefont {Milovanovi{\'{c}}},\ and\ \citenamefont {Peeters}}]{ref14}%
  \BibitemOpen
  \bibfield  {author} {\bibinfo {author} {\bibfnamefont {R.~V.}\ \bibnamefont {Pottelberge}}, \bibinfo {author} {\bibfnamefont {D.}~\bibnamefont {Moldovan}}, \bibinfo {author} {\bibfnamefont {S.~P.}\ \bibnamefont {Milovanovi{\'{c}}}},\ and\ \bibinfo {author} {\bibfnamefont {F.~M.}\ \bibnamefont {Peeters}},\ }\href {https://doi.org/10.1088/2053-1583/ab3feb} {\bibfield  {journal} {\bibinfo  {journal} {2D Mater.}\ }\textbf {\bibinfo {volume} {6}},\ \bibinfo {pages} {045047} (\bibinfo {year} {2019})}\BibitemShut {NoStop}%
\bibitem [{\citenamefont {Wang}\ \emph {et~al.}(2020)\citenamefont {Wang}, \citenamefont {Andelkovic}, \citenamefont {Wang},\ and\ \citenamefont {Peeters}}]{ref15}%
  \BibitemOpen
  \bibfield  {author} {\bibinfo {author} {\bibfnamefont {J.}~\bibnamefont {Wang}}, \bibinfo {author} {\bibfnamefont {M.}~\bibnamefont {Andelkovic}}, \bibinfo {author} {\bibfnamefont {G.}~\bibnamefont {Wang}},\ and\ \bibinfo {author} {\bibfnamefont {F.~M.}\ \bibnamefont {Peeters}},\ }\href {https://doi.org/10.1103/PhysRevB.102.064108} {\bibfield  {journal} {\bibinfo  {journal} {Phys. Rev. B}\ }\textbf {\bibinfo {volume} {102}},\ \bibinfo {pages} {064108} (\bibinfo {year} {2020})}\BibitemShut {NoStop}%
\bibitem [{\citenamefont {Fogler}\ \emph {et~al.}(2007)\citenamefont {Fogler}, \citenamefont {Novikov},\ and\ \citenamefont {Shklovskii}}]{ref16}%
  \BibitemOpen
  \bibfield  {author} {\bibinfo {author} {\bibfnamefont {M.~M.}\ \bibnamefont {Fogler}}, \bibinfo {author} {\bibfnamefont {D.~S.}\ \bibnamefont {Novikov}},\ and\ \bibinfo {author} {\bibfnamefont {B.~I.}\ \bibnamefont {Shklovskii}},\ }\href {https://doi.org/10.1103/PhysRevB.76.233402} {\bibfield  {journal} {\bibinfo  {journal} {Phys. Rev. B}\ }\textbf {\bibinfo {volume} {76}},\ \bibinfo {pages} {233402} (\bibinfo {year} {2007})}\BibitemShut {NoStop}%
\bibitem [{\citenamefont {Terekhov}\ \emph {et~al.}(2008)\citenamefont {Terekhov}, \citenamefont {Milstein}, \citenamefont {Kotov},\ and\ \citenamefont {Sushkov}}]{ref17}%
  \BibitemOpen
  \bibfield  {author} {\bibinfo {author} {\bibfnamefont {I.~S.}\ \bibnamefont {Terekhov}}, \bibinfo {author} {\bibfnamefont {A.~I.}\ \bibnamefont {Milstein}}, \bibinfo {author} {\bibfnamefont {V.~N.}\ \bibnamefont {Kotov}},\ and\ \bibinfo {author} {\bibfnamefont {O.~P.}\ \bibnamefont {Sushkov}},\ }\href {https://doi.org/10.1103/PhysRevLett.100.076803} {\bibfield  {journal} {\bibinfo  {journal} {Phys. Rev. Lett.}\ }\textbf {\bibinfo {volume} {100}},\ \bibinfo {pages} {076803} (\bibinfo {year} {2008})}\BibitemShut {NoStop}%
\bibitem [{\citenamefont {Neto}\ \emph {et~al.}(2009)\citenamefont {Neto}, \citenamefont {Kotov}, \citenamefont {Nilsson}, \citenamefont {Pereira}, \citenamefont {Peres},\ and\ \citenamefont {Uchoa}}]{ref18}%
  \BibitemOpen
  \bibfield  {author} {\bibinfo {author} {\bibfnamefont {A.~C.}\ \bibnamefont {Neto}}, \bibinfo {author} {\bibfnamefont {V.}~\bibnamefont {Kotov}}, \bibinfo {author} {\bibfnamefont {J.}~\bibnamefont {Nilsson}}, \bibinfo {author} {\bibfnamefont {V.}~\bibnamefont {Pereira}}, \bibinfo {author} {\bibfnamefont {N.}~\bibnamefont {Peres}},\ and\ \bibinfo {author} {\bibfnamefont {B.}~\bibnamefont {Uchoa}},\ }\href {https://doi.org/10.1016/j.ssc.2009.02.040} {\bibfield  {journal} {\bibinfo  {journal} {Solid State Commun.}\ }\textbf {\bibinfo {volume} {149}},\ \bibinfo {pages} {1094} (\bibinfo {year} {2009})}\BibitemShut {NoStop}%
\bibitem [{\citenamefont {Novikov}(2007)}]{ref19}%
  \BibitemOpen
  \bibfield  {author} {\bibinfo {author} {\bibfnamefont {D.~S.}\ \bibnamefont {Novikov}},\ }\href {https://doi.org/10.1103/PhysRevB.76.245435} {\bibfield  {journal} {\bibinfo  {journal} {Phys. Rev. B}\ }\textbf {\bibinfo {volume} {76}},\ \bibinfo {pages} {245435} (\bibinfo {year} {2007})}\BibitemShut {NoStop}%
\bibitem [{\citenamefont {Kotov}\ \emph {et~al.}(2012)\citenamefont {Kotov}, \citenamefont {Uchoa}, \citenamefont {Pereira}, \citenamefont {Guinea},\ and\ \citenamefont {Castro~Neto}}]{ref20}%
  \BibitemOpen
  \bibfield  {author} {\bibinfo {author} {\bibfnamefont {V.~N.}\ \bibnamefont {Kotov}}, \bibinfo {author} {\bibfnamefont {B.}~\bibnamefont {Uchoa}}, \bibinfo {author} {\bibfnamefont {V.~M.}\ \bibnamefont {Pereira}}, \bibinfo {author} {\bibfnamefont {F.}~\bibnamefont {Guinea}},\ and\ \bibinfo {author} {\bibfnamefont {A.~H.}\ \bibnamefont {Castro~Neto}},\ }\href {https://doi.org/10.1103/RevModPhys.84.1067} {\bibfield  {journal} {\bibinfo  {journal} {Rev. Mod. Phys.}\ }\textbf {\bibinfo {volume} {84}},\ \bibinfo {pages} {1067} (\bibinfo {year} {2012})}\BibitemShut {NoStop}%
\bibitem [{\citenamefont {Moldovan}\ \emph {et~al.}(2017{\natexlab{a}})\citenamefont {Moldovan}, \citenamefont {Masir},\ and\ \citenamefont {Peeters}}]{ref21}%
  \BibitemOpen
  \bibfield  {author} {\bibinfo {author} {\bibfnamefont {D.}~\bibnamefont {Moldovan}}, \bibinfo {author} {\bibfnamefont {M.~R.}\ \bibnamefont {Masir}},\ and\ \bibinfo {author} {\bibfnamefont {F.~M.}\ \bibnamefont {Peeters}},\ }\href {https://doi.org/10.1088/2053-1583/aa9647} {\bibfield  {journal} {\bibinfo  {journal} {2D Mater.}\ }\textbf {\bibinfo {volume} {5}},\ \bibinfo {pages} {015017} (\bibinfo {year} {2017}{\natexlab{a}})}\BibitemShut {NoStop}%
\bibitem [{\citenamefont {De~Martino}\ \emph {et~al.}(2014)\citenamefont {De~Martino}, \citenamefont {Kl\"opfer}, \citenamefont {Matrasulov},\ and\ \citenamefont {Egger}}]{ref22}%
  \BibitemOpen
  \bibfield  {author} {\bibinfo {author} {\bibfnamefont {A.}~\bibnamefont {De~Martino}}, \bibinfo {author} {\bibfnamefont {D.}~\bibnamefont {Kl\"opfer}}, \bibinfo {author} {\bibfnamefont {D.}~\bibnamefont {Matrasulov}},\ and\ \bibinfo {author} {\bibfnamefont {R.}~\bibnamefont {Egger}},\ }\href {https://doi.org/10.1103/PhysRevLett.112.186603} {\bibfield  {journal} {\bibinfo  {journal} {Phys. Rev. Lett.}\ }\textbf {\bibinfo {volume} {112}},\ \bibinfo {pages} {186603} (\bibinfo {year} {2014})}\BibitemShut {NoStop}%
\bibitem [{\citenamefont {Kl\"{o}pfer}\ \emph {et~al.}(2014)\citenamefont {Kl\"{o}pfer}, \citenamefont {Martino}, \citenamefont {Matrasulov},\ and\ \citenamefont {Egger}}]{ref23}%
  \BibitemOpen
  \bibfield  {author} {\bibinfo {author} {\bibfnamefont {D.}~\bibnamefont {Kl\"{o}pfer}}, \bibinfo {author} {\bibfnamefont {A.~D.}\ \bibnamefont {Martino}}, \bibinfo {author} {\bibfnamefont {D.~U.}\ \bibnamefont {Matrasulov}},\ and\ \bibinfo {author} {\bibfnamefont {R.}~\bibnamefont {Egger}},\ }\href {https://doi.org/10.1140/epjb/e2014-50414-8} {\bibfield  {journal} {\bibinfo  {journal} {Eur. Phys. J. B}\ }\textbf {\bibinfo {volume} {87}},\ \bibinfo {pages} {187} (\bibinfo {year} {2014})}\BibitemShut {NoStop}%
\bibitem [{\citenamefont {Gorbar}\ \emph {et~al.}(2015)\citenamefont {Gorbar}, \citenamefont {Gusynin},\ and\ \citenamefont {Sobol}}]{ref24}%
  \BibitemOpen
  \bibfield  {author} {\bibinfo {author} {\bibfnamefont {E.~V.}\ \bibnamefont {Gorbar}}, \bibinfo {author} {\bibfnamefont {V.~P.}\ \bibnamefont {Gusynin}},\ and\ \bibinfo {author} {\bibfnamefont {O.~O.}\ \bibnamefont {Sobol}},\ }\href {https://doi.org/10.1103/PhysRevB.92.235417} {\bibfield  {journal} {\bibinfo  {journal} {Phys. Rev. B}\ }\textbf {\bibinfo {volume} {92}},\ \bibinfo {pages} {235417} (\bibinfo {year} {2015})}\BibitemShut {NoStop}%
\bibitem [{\citenamefont {Van~Pottelberge}\ \emph {et~al.}(2018)\citenamefont {Van~Pottelberge}, \citenamefont {Van~Duppen},\ and\ \citenamefont {Peeters}}]{ref25}%
  \BibitemOpen
  \bibfield  {author} {\bibinfo {author} {\bibfnamefont {R.}~\bibnamefont {Van~Pottelberge}}, \bibinfo {author} {\bibfnamefont {B.}~\bibnamefont {Van~Duppen}},\ and\ \bibinfo {author} {\bibfnamefont {F.~M.}\ \bibnamefont {Peeters}},\ }\href {https://doi.org/10.1103/PhysRevB.98.165420} {\bibfield  {journal} {\bibinfo  {journal} {Phys. Rev. B}\ }\textbf {\bibinfo {volume} {98}},\ \bibinfo {pages} {165420} (\bibinfo {year} {2018})}\BibitemShut {NoStop}%
\bibitem [{\citenamefont {Lu}\ \emph {et~al.}(2019{\natexlab{b}})\citenamefont {Lu}, \citenamefont {Tsai}, \citenamefont {Tatan}, \citenamefont {Wickenburg}, \citenamefont {Omrani}, \citenamefont {Wong}, \citenamefont {Riss}, \citenamefont {Piatti}, \citenamefont {Watanabe}, \citenamefont {Taniguchi}, \citenamefont {Zettl}, \citenamefont {Pereira},\ and\ \citenamefont {Crommie}}]{ref26}%
  \BibitemOpen
  \bibfield  {author} {\bibinfo {author} {\bibfnamefont {J.}~\bibnamefont {Lu}}, \bibinfo {author} {\bibfnamefont {H.-Z.}\ \bibnamefont {Tsai}}, \bibinfo {author} {\bibfnamefont {A.~N.}\ \bibnamefont {Tatan}}, \bibinfo {author} {\bibfnamefont {S.}~\bibnamefont {Wickenburg}}, \bibinfo {author} {\bibfnamefont {A.~A.}\ \bibnamefont {Omrani}}, \bibinfo {author} {\bibfnamefont {D.}~\bibnamefont {Wong}}, \bibinfo {author} {\bibfnamefont {A.}~\bibnamefont {Riss}}, \bibinfo {author} {\bibfnamefont {E.}~\bibnamefont {Piatti}}, \bibinfo {author} {\bibfnamefont {K.}~\bibnamefont {Watanabe}}, \bibinfo {author} {\bibfnamefont {T.}~\bibnamefont {Taniguchi}}, \bibinfo {author} {\bibfnamefont {A.}~\bibnamefont {Zettl}}, \bibinfo {author} {\bibfnamefont {V.~M.}\ \bibnamefont {Pereira}},\ and\ \bibinfo {author} {\bibfnamefont {M.~F.}\ \bibnamefont {Crommie}},\ }\href {https://doi.org/10.1038/s41467-019-08371-2} {\bibfield  {journal} {\bibinfo  {journal} {Nat. Commun.}\ }\textbf {\bibinfo {volume} {10}},\ \bibinfo {pages} {477}
  (\bibinfo {year} {2019}{\natexlab{b}})}\BibitemShut {NoStop}%
\bibitem [{\citenamefont {Krainov}\ and\ \citenamefont {Zakharov}(1973)}]{ref27}%
  \BibitemOpen
  \bibfield  {author} {\bibinfo {author} {\bibfnamefont {V.~P.}\ \bibnamefont {Krainov}}\ and\ \bibinfo {author} {\bibfnamefont {S.~I.}\ \bibnamefont {Zakharov}},\ }\href {https://www.osti.gov/biblio/4445971} {\bibfield  {journal} {\bibinfo  {journal} {Sov. Phys.-JETP}\ }\textbf {\bibinfo {volume} {37}},\ \bibinfo {pages} {983} (\bibinfo {year} {1973})}\BibitemShut {NoStop}%
\bibitem [{\citenamefont {Oraevskii}\ \emph {et~al.}(1977)\citenamefont {Oraevskii}, \citenamefont {Rex},\ and\ \citenamefont {Semikoz}}]{ref28}%
  \BibitemOpen
  \bibfield  {author} {\bibinfo {author} {\bibfnamefont {V.~N.}\ \bibnamefont {Oraevskii}}, \bibinfo {author} {\bibfnamefont {A.~I.}\ \bibnamefont {Rex}},\ and\ \bibinfo {author} {\bibfnamefont {V.~B.}\ \bibnamefont {Semikoz}},\ }\href {http://www.jetp.ras.ru/cgi-bin/dn/e_045_03_0428.pdf} {\bibfield  {journal} {\bibinfo  {journal} {Sov. Phys.-JETP}\ }\textbf {\bibinfo {volume} {45}},\ \bibinfo {pages} {428} (\bibinfo {year} {1977})}\BibitemShut {NoStop}%
\bibitem [{\citenamefont {Karnakov}\ and\ \citenamefont {Popov}(2003)}]{ref29}%
  \BibitemOpen
  \bibfield  {author} {\bibinfo {author} {\bibfnamefont {B.}~\bibnamefont {Karnakov}}\ and\ \bibinfo {author} {\bibfnamefont {V.}~\bibnamefont {Popov}},\ }\href {https://doi.org/10.1134/1.1633946} {\bibfield  {journal} {\bibinfo  {journal} {J. Exp. Theor. Phys.}\ }\textbf {\bibinfo {volume} {97}},\ \bibinfo {pages} {890–914} (\bibinfo {year} {2003})}\BibitemShut {NoStop}%
\bibitem [{\citenamefont {Vysotskii}\ and\ \citenamefont {Godunov}(2014)}]{ref30}%
  \BibitemOpen
  \bibfield  {author} {\bibinfo {author} {\bibfnamefont {M.~I.}\ \bibnamefont {Vysotskii}}\ and\ \bibinfo {author} {\bibfnamefont {S.~I.}\ \bibnamefont {Godunov}},\ }\href {https://doi.org/10.3367/ufne.0184.201402j.0206} {\bibfield  {journal} {\bibinfo  {journal} {Physics-Uspekhi}\ }\textbf {\bibinfo {volume} {57}},\ \bibinfo {pages} {194} (\bibinfo {year} {2014})}\BibitemShut {NoStop}%
\bibitem [{\citenamefont {Gorbar}\ \emph {et~al.}(2018)\citenamefont {Gorbar}, \citenamefont {Gusynin},\ and\ \citenamefont {Sobol}}]{ref31}%
  \BibitemOpen
  \bibfield  {author} {\bibinfo {author} {\bibfnamefont {E.~V.}\ \bibnamefont {Gorbar}}, \bibinfo {author} {\bibfnamefont {V.~P.}\ \bibnamefont {Gusynin}},\ and\ \bibinfo {author} {\bibfnamefont {O.}~\bibnamefont {Sobol}},\ }\href {https://doi.org/10.1063/1.5034149} {\bibfield  {journal} {\bibinfo  {journal} {Low Temperature Physics}\ }\textbf {\bibinfo {volume} {44}},\ \bibinfo {pages} {371} (\bibinfo {year} {2018})}\BibitemShut {NoStop}%
\bibitem [{\citenamefont {Gamayun}\ \emph {et~al.}(2011)\citenamefont {Gamayun}, \citenamefont {Gorbar},\ and\ \citenamefont {Gusynin}}]{ref32}%
  \BibitemOpen
  \bibfield  {author} {\bibinfo {author} {\bibfnamefont {O.~V.}\ \bibnamefont {Gamayun}}, \bibinfo {author} {\bibfnamefont {E.~V.}\ \bibnamefont {Gorbar}},\ and\ \bibinfo {author} {\bibfnamefont {V.~P.}\ \bibnamefont {Gusynin}},\ }\href {https://doi.org/10.1103/PhysRevB.83.235104} {\bibfield  {journal} {\bibinfo  {journal} {Phys. Rev. B}\ }\textbf {\bibinfo {volume} {83}},\ \bibinfo {pages} {235104} (\bibinfo {year} {2011})}\BibitemShut {NoStop}%
\bibitem [{\citenamefont {Valenzuela}\ \emph {et~al.}(2016)\citenamefont {Valenzuela}, \citenamefont {Hern{\'{a}}ndez-Ortiz}, \citenamefont {Loewe},\ and\ \citenamefont {Raya}}]{ref33}%
  \BibitemOpen
  \bibfield  {author} {\bibinfo {author} {\bibfnamefont {D.}~\bibnamefont {Valenzuela}}, \bibinfo {author} {\bibfnamefont {S.}~\bibnamefont {Hern{\'{a}}ndez-Ortiz}}, \bibinfo {author} {\bibfnamefont {M.}~\bibnamefont {Loewe}},\ and\ \bibinfo {author} {\bibfnamefont {A.}~\bibnamefont {Raya}},\ }\href {https://doi.org/10.1088/1751-8113/49/49/495302} {\bibfield  {journal} {\bibinfo  {journal} {J. Phys. A: Math. Theor.}\ }\textbf {\bibinfo {volume} {49}},\ \bibinfo {pages} {495302} (\bibinfo {year} {2016})}\BibitemShut {NoStop}%
\bibitem [{\citenamefont {Zhang}\ \emph {et~al.}(2012)\citenamefont {Zhang}, \citenamefont {Barlas},\ and\ \citenamefont {Yang}}]{ref34}%
  \BibitemOpen
  \bibfield  {author} {\bibinfo {author} {\bibfnamefont {Y.}~\bibnamefont {Zhang}}, \bibinfo {author} {\bibfnamefont {Y.}~\bibnamefont {Barlas}},\ and\ \bibinfo {author} {\bibfnamefont {K.}~\bibnamefont {Yang}},\ }\href {https://doi.org/10.1103/PhysRevB.85.165423} {\bibfield  {journal} {\bibinfo  {journal} {Phys. Rev. B}\ }\textbf {\bibinfo {volume} {85}},\ \bibinfo {pages} {165423} (\bibinfo {year} {2012})}\BibitemShut {NoStop}%
\bibitem [{\citenamefont {Maiera}\ and\ \citenamefont {Siedentopb}(2012)}]{ref35}%
  \BibitemOpen
  \bibfield  {author} {\bibinfo {author} {\bibfnamefont {T.}~\bibnamefont {Maiera}}\ and\ \bibinfo {author} {\bibfnamefont {H.}~\bibnamefont {Siedentopb}},\ }\href {https://doi.org/10.1063/1.4728982} {\bibfield  {journal} {\bibinfo  {journal} {J. Math. Phys.}\ }\textbf {\bibinfo {volume} {53}},\ \bibinfo {pages} {095207} (\bibinfo {year} {2012})}\BibitemShut {NoStop}%
\bibitem [{\citenamefont {Kim}\ and\ \citenamefont {{Eric Yang}}(2014)}]{ref36}%
  \BibitemOpen
  \bibfield  {author} {\bibinfo {author} {\bibfnamefont {S.}~\bibnamefont {Kim}}\ and\ \bibinfo {author} {\bibfnamefont {S.-R.}\ \bibnamefont {{Eric Yang}}},\ }\href {https://doi.org/https://doi.org/10.1016/j.aop.2014.04.022} {\bibfield  {journal} {\bibinfo  {journal} {Ann. Phys.}\ }\textbf {\bibinfo {volume} {347}},\ \bibinfo {pages} {21} (\bibinfo {year} {2014})}\BibitemShut {NoStop}%
\bibitem [{\citenamefont {Eren}\ and\ \citenamefont {Güçlü}(2022)}]{ref37}%
  \BibitemOpen
  \bibfield  {author} {\bibinfo {author} {\bibfnamefont {I.}~\bibnamefont {Eren}}\ and\ \bibinfo {author} {\bibfnamefont {A.}~\bibnamefont {Güçlü}},\ }\href {https://doi.org/https://doi.org/10.1016/j.ssc.2022.114763} {\bibfield  {journal} {\bibinfo  {journal} {Solid State Communications}\ }\textbf {\bibinfo {volume} {351}},\ \bibinfo {pages} {114763} (\bibinfo {year} {2022})}\BibitemShut {NoStop}%
\bibitem [{\citenamefont {Pereira}\ \emph {et~al.}(2008)\citenamefont {Pereira}, \citenamefont {Kotov},\ and\ \citenamefont {Castro~Neto}}]{ref38}%
  \BibitemOpen
  \bibfield  {author} {\bibinfo {author} {\bibfnamefont {V.~M.}\ \bibnamefont {Pereira}}, \bibinfo {author} {\bibfnamefont {V.~N.}\ \bibnamefont {Kotov}},\ and\ \bibinfo {author} {\bibfnamefont {A.~H.}\ \bibnamefont {Castro~Neto}},\ }\href {https://doi.org/10.1103/PhysRevB.78.085101} {\bibfield  {journal} {\bibinfo  {journal} {Phys. Rev. B}\ }\textbf {\bibinfo {volume} {78}},\ \bibinfo {pages} {085101} (\bibinfo {year} {2008})}\BibitemShut {NoStop}%
\bibitem [{\citenamefont {Zhu}\ \emph {et~al.}(2009)\citenamefont {Zhu}, \citenamefont {Wang}, \citenamefont {Shi}, \citenamefont {Szeto}, \citenamefont {Chen},\ and\ \citenamefont {Hou}}]{ref39}%
  \BibitemOpen
  \bibfield  {author} {\bibinfo {author} {\bibfnamefont {W.}~\bibnamefont {Zhu}}, \bibinfo {author} {\bibfnamefont {Z.}~\bibnamefont {Wang}}, \bibinfo {author} {\bibfnamefont {Q.}~\bibnamefont {Shi}}, \bibinfo {author} {\bibfnamefont {K.~Y.}\ \bibnamefont {Szeto}}, \bibinfo {author} {\bibfnamefont {J.}~\bibnamefont {Chen}},\ and\ \bibinfo {author} {\bibfnamefont {J.~G.}\ \bibnamefont {Hou}},\ }\href {https://doi.org/10.1103/PhysRevB.79.155430} {\bibfield  {journal} {\bibinfo  {journal} {Phys. Rev. B}\ }\textbf {\bibinfo {volume} {79}},\ \bibinfo {pages} {155430} (\bibinfo {year} {2009})}\BibitemShut {NoStop}%
\bibitem [{\citenamefont {Zhu}\ \emph {et~al.}(2014)\citenamefont {Zhu}, \citenamefont {Liu},\ and\ \citenamefont {Yang}}]{ref40}%
  \BibitemOpen
  \bibfield  {author} {\bibinfo {author} {\bibfnamefont {J.-L.}\ \bibnamefont {Zhu}}, \bibinfo {author} {\bibfnamefont {C.}~\bibnamefont {Liu}},\ and\ \bibinfo {author} {\bibfnamefont {N.}~\bibnamefont {Yang}},\ }\href {https://doi.org/10.1103/PhysRevB.90.125405} {\bibfield  {journal} {\bibinfo  {journal} {Phys. Rev. B}\ }\textbf {\bibinfo {volume} {90}},\ \bibinfo {pages} {125405} (\bibinfo {year} {2014})}\BibitemShut {NoStop}%
\bibitem [{\citenamefont {Giovannetti}\ \emph {et~al.}(2007)\citenamefont {Giovannetti}, \citenamefont {Khomyakov}, \citenamefont {Brocks}, \citenamefont {Kelly},\ and\ \citenamefont {van~den Brink}}]{ref41}%
  \BibitemOpen
  \bibfield  {author} {\bibinfo {author} {\bibfnamefont {G.}~\bibnamefont {Giovannetti}}, \bibinfo {author} {\bibfnamefont {P.~A.}\ \bibnamefont {Khomyakov}}, \bibinfo {author} {\bibfnamefont {G.}~\bibnamefont {Brocks}}, \bibinfo {author} {\bibfnamefont {P.~J.}\ \bibnamefont {Kelly}},\ and\ \bibinfo {author} {\bibfnamefont {J.}~\bibnamefont {van~den Brink}},\ }\href {https://doi.org/10.1103/PhysRevB.76.073103} {\bibfield  {journal} {\bibinfo  {journal} {Phys. Rev. B}\ }\textbf {\bibinfo {volume} {76}},\ \bibinfo {pages} {073103} (\bibinfo {year} {2007})}\BibitemShut {NoStop}%
\bibitem [{\citenamefont {Zhou}\ \emph {et~al.}(2007)\citenamefont {Zhou}, \citenamefont {Gweon}, \citenamefont {Fedorov}, \citenamefont {First}, \citenamefont {de~Heer}, \citenamefont {Lee}, \citenamefont {Guinea}, \citenamefont {H.},\ and\ \citenamefont {A.}}]{ref42}%
  \BibitemOpen
  \bibfield  {author} {\bibinfo {author} {\bibfnamefont {S.~Y.}\ \bibnamefont {Zhou}}, \bibinfo {author} {\bibfnamefont {G.~H.}\ \bibnamefont {Gweon}}, \bibinfo {author} {\bibfnamefont {A.~V.}\ \bibnamefont {Fedorov}}, \bibinfo {author} {\bibfnamefont {P.~N.}\ \bibnamefont {First}}, \bibinfo {author} {\bibfnamefont {W.~A.}\ \bibnamefont {de~Heer}}, \bibinfo {author} {\bibfnamefont {D.~H.}\ \bibnamefont {Lee}}, \bibinfo {author} {\bibfnamefont {F.}~\bibnamefont {Guinea}}, \bibinfo {author} {\bibfnamefont {C.~N.~A.}\ \bibnamefont {H.}},\ and\ \bibinfo {author} {\bibfnamefont {L.}~\bibnamefont {A.}},\ }\href {https://doi.org/10.1038/nmat2003} {\bibfield  {journal} {\bibinfo  {journal} {Nature Materials}\ }\textbf {\bibinfo {volume} {6}},\ \bibinfo {pages} {6} (\bibinfo {year} {2007})}\BibitemShut {NoStop}%
\bibitem [{\citenamefont {Popov}(1971)}]{ref43}%
  \BibitemOpen
  \bibfield  {author} {\bibinfo {author} {\bibfnamefont {V.~S.}\ \bibnamefont {Popov}},\ }\href {http://jetp.ras.ru/cgi-bin/dn/e_033_04_0665.pdf} {\bibfield  {journal} {\bibinfo  {journal} {Sov. J. Nucl. Phys.}\ }\textbf {\bibinfo {volume} {12}} (\bibinfo {year} {1971})}\BibitemShut {NoStop}%
\bibitem [{\citenamefont {Moldovan}\ \emph {et~al.}(2017{\natexlab{b}})\citenamefont {Moldovan}, \citenamefont {Andelkovic},\ and\ \citenamefont {Peeters}}]{ref44}%
  \BibitemOpen
  \bibfield  {author} {\bibinfo {author} {\bibfnamefont {D.}~\bibnamefont {Moldovan}}, \bibinfo {author} {\bibfnamefont {M.}~\bibnamefont {Andelkovic}},\ and\ \bibinfo {author} {\bibfnamefont {F.~M.}\ \bibnamefont {Peeters}},\ }\href {https://doi.org/10.5281/zenodo.826942} {\bibinfo {title} {Pybinding v0.9.4: A python package for tight-binding calculations}} (\bibinfo {year} {2017}{\natexlab{b}})\BibitemShut {NoStop}%
\bibitem [{\citenamefont {Khalilov}\ and\ \citenamefont {Ho}(1998)}]{ref45}%
  \BibitemOpen
  \bibfield  {author} {\bibinfo {author} {\bibfnamefont {V.~R.}\ \bibnamefont {Khalilov}}\ and\ \bibinfo {author} {\bibfnamefont {C.~L.}\ \bibnamefont {Ho}},\ }\href {https://doi.org/10.1142/S0217732398000668} {\bibfield  {journal} {\bibinfo  {journal} {Mod. Phys. Lett. A}\ }\textbf {\bibinfo {volume} {13}} (\bibinfo {year} {1998})}\BibitemShut {NoStop}%
\end{thebibliography}%
\end{document}